\documentclass[aps,prl,twocolumn,showpacs,superscriptaddress,groupedaddress]{revtex4} 
\usepackage{graphicx}  
\usepackage{dcolumn}   
\usepackage{bm}        
\usepackage{amssymb}
\usepackage{float}
\usepackage{amsmath}
\usepackage{color}
\usepackage{epsfig}

\begin{document}

\title{Low power sessile droplets actuation via modulated surface acoustic waves}

\date{\today}

\begin{abstract}

Low power actuation of sessile droplets is of primary interest for portable or hybrid lab-on-a-chip and harmless manipulation of biofluids. In this paper, we show that the acoustic power required to move or deform droplets via surface acoustic waves can be substantially reduced through the forcing of the drops inertio-capillary modes of vibrations. Indeed, harmonic, superharmonic and subharmonic (parametric) excitation of these modes are observed when the high frequency acoustic signal (19.5 MHz) is modulated around Rayleigh-Lamb inertio-capillary  frequencies. This resonant behavior results in larger oscillations and quicker motion of the drops than in the non-modulated case.

\end{abstract}

\author{Michael Baudoin}
\email{michael.baudoin@univ-lille1.fr}
\affiliation{International Associated Laboratory LEMAC, IEMN, UMR CNRS 8520, Universit\'{e} des Sciences et Technologies de Lille and EC Lille, 59652 Villeneuve dAscq C\'{e}dex, France}
\author{Philippe Brunet}
\affiliation{Laboratoire MSC, UMR 75057, 10 rue Alice Domon et L\'{e}onie Duquet, 75205 Paris, France}
\author{Olivier Bou Matar}
\affiliation{International Associated Laboratory LEMAC, IEMN, UMR CNRS 8520, Universit\'{e} des Sciences et Technologies de Lille and EC Lille, 59652 Villeneuve dAscq C\'{e}dex, France}
\author{Etienne Herth}
\affiliation{International Associated Laboratory LEMAC, IEMN, UMR CNRS 8520, Universit\'{e} des Sciences et Technologies de Lille and EC Lille, 59652 Villeneuve dAscq C\'{e}dex, France}

\maketitle

\textit{The following article has been submitted to Applied Physics Letters (\url{http://apl.aip.org/})}

One of the challenges in droplet microfluidics is to overcome surface capillary forces and contact line retention forces, which prevent the motion and deformation of the drop \cite{arfm_stone_2004}. Different techniques such as electrowetting \cite{apl_pollack_2000}, optical toolbox \cite{loc_baroud_2007} based on thermocapillary forces induced by a focused laser, or ultrasonic surface acoustic waves (SAW) have been developed to perform simple operations on droplets. In particular, SAW are efficient to achieve actuation, atomization, jetting, oscillations, or mixing of small quantities of liquid, either lying on a solid substrate or entrapped in confined geometries \cite{rmp_friend_2011}. However, due to nonlinear coupling between thermal and acoustical mode, SAW can induce quick temperature increase in fluid samples \cite{kondoh_saa_2009,phd_beyssen_2006}. This can be detrimental for the manipulation of biofluids (albumin coagulates in a few seconds when excited by SAW), or for hybrid SPR (Surface Plasmon Resonance)/SAW lab-on-a chip \cite{loc_renaudin_2010} (substrate heating causes a shift in SPR reflectivity). To enhance the droplet response, particular attention has been devoted to the design of the actuators \cite{prl_tan_2009} and to the chemical treatment of the surface \cite{sm_wixforth_2003} to reduce hysteresis and modify the contact angle. However less effort has been dedicated to the optimization of the acoustic signal in relation to the natural frequencies of the drop. Rayleigh in 1879 \cite{prsl_rayleigh_1879} and Lamb in 1932 \cite{cup_lamb_1932} have identified oscillation modes resulting from a competition between inertia and surface tension. While these have been first described for levitating drops, they can also be adapted to sessile drops. In this case, the droplet vibration is affected by the wettability of the surface \cite{jfm_strani_1984} and the pinning of the contact line \cite{epje_noblin_2004}. 

These low-frequency oscillations (typically from 10 to 200 Hz for millimeter-sized drops) are observed when a drop lying on a solid substrate is subjected to sinusoidal SAW of much higher frequency (about 20 MHz) \cite{jjap_yamakita_1999,pre_brunet_2010}. The free surface deformation results from nonlinear acoustic forces (acoustic radiation pressure and acoustic streaming bulk force), while the detailed mechanism of excitation of these modes has not been elucidated yet. The acoustic radiation pressure induces a stress at the surface of the drop, while acoustic streaming induces internal flow, which can also contribute to the droplet deformation. The respective magnitudes of these two force fields depend on the properties of the acoustic field inside the drop and thus on the frequency of excitation, the size of the drop, and the acoustic attenuation length \cite{jsv_alzuaga_2005,pre_brunet_2010}.  

In this paper, we show that modulations of the acoustic signal around Rayleigh-Lamb characteristic frequency and twice this frequency, result respectively in harmonic and parametric response. This latter was predicted theoretically by Papoular and Parayre for levitating drops \cite{prl_papoular_1997} but not observed experimentally. This resonant behavior leads to higher amplitude drop oscillations compared to the non-modulated case at same input acoustic power. We also show that these oscillations promote droplets mobility.
\begin{figure}[htbp]
\includegraphics[width=0.5\textwidth]{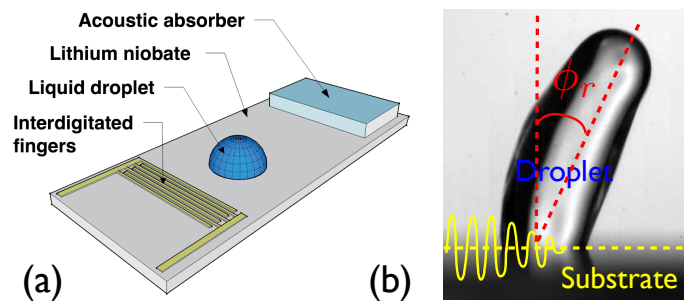}\hfill 
\caption{(color online) (a) Sketch of the experimental setup. (b) Drop undergoing large deformation along the acoustic wave refraction angle $\phi_r$. \label{fig:idt}}
\end{figure}

SAW are generated at the surface of a 1.05 mm thick piezoelectric substrate (X-cut Niobate Lithium LiNbO$_3$) by a transducer consisting of interdigitated fingers, Fig. \ref{fig:idt}. These fingers are designed with the following process: 1) A titanium (Ti) layer of 20 nm and a gold (Au) layer of 200 nm are successively sputtered on a LiNbO$_3$ substrate 2) The substrate is coated with AZnlof2020 resist, which is patterned by conventional photolithography technique 3) The Au/Ti layers are successively wet-etched by potassium iodide (KI) and hydrofluoric acid (HF 50\%), and 4) AZnlof2020 is removed by acetone. The width of the fingers and their distance are both equal to 43.75 $\mu$m, leading to a characteristic frequency of 19.5 MHz, which is used as the carrier frequency $f_c$. A periodic sinusoidal voltage is applied at this frequency with a high frequency generator (IFR 2023A) and amplified with a home made amplifier. This carrier signal is modulated by a square wave switching between 1 and 0, at frequency $f_m \ll f_c$. The amplitude $d$ of the SAW is measured with a Mach-Zender laser interferometer (BMI-SH130). The surface of the substrate is treated with hydrophobic coating (monolayer of OTS) leading to advancing and receding contact angles of  $\theta_a = 108^o$ and $\theta_r = 99^o$ respectively, measured with a  Kruss DSA100 goniometer. A 7.5 $\mu$l droplet of water is then placed on the substrate (a sensitivity analysis of the droplet response according to its volume has been conducted in Ref.~\onlinecite{pre_brunet_2010}). The droplet dynamics is observed via a high speed camera (Photron SA3) and recorded at 2000 frames per second. To avoid pollution of the surface by impurities and to obtain reproducible results, all the experiments have been carried out in our laboratory class 1000 clean room. 
\begin{figure}[htbp]
\includegraphics[width=0.45 \textwidth]{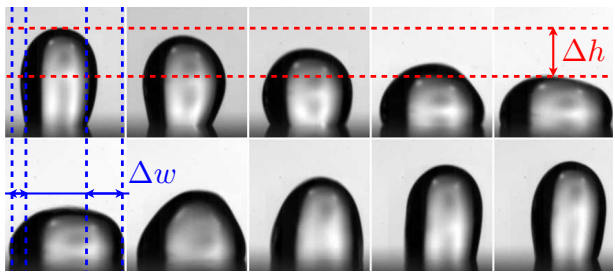}
\caption{(color online) Dipolar (degree $l$=2) zonal oscillations of a drop of 7.5 $\mu$l subjected to a SAW of carrier frequency $f_c$ = 19.5 MHz, modulation frequency $f_m$ = 52.5 Hz, and amplitude $d$ = 1.38 nm. The time elapsed between two successive snapshots is 2 ms. $\Delta h$ and $\Delta w$ are respectively the longitudinal and lateral amplitude of oscillation. \label{fig:montage}}
\end{figure}
\begin{figure}[htbp]
\includegraphics[width=0.5\textwidth]{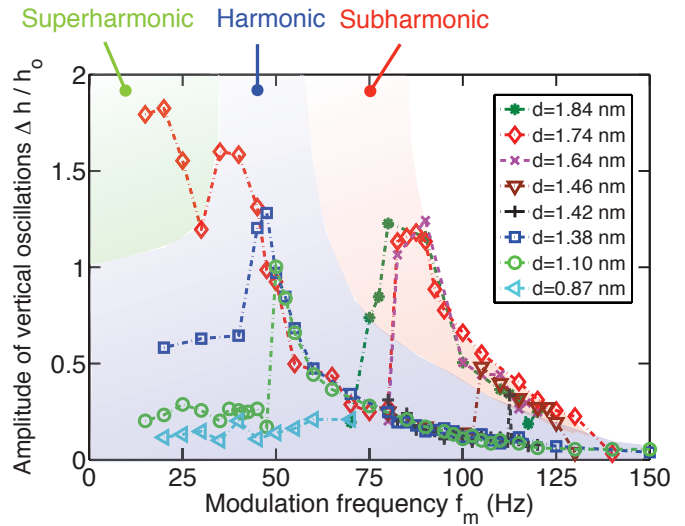}
\caption{(color online) Amplitude of vertical oscillations of the drop $\Delta h$ divided by the initial height of the droplet $h_o$ as a function of the modulation frequency $f_m$ for different amplitudes $d$ of the surface acoustic wave. In the green, blue and red region, the droplet response is respectively superharmonic, harmonic and subharmonic (compared to the frequency of modulation). \label{fig:amplitudepcomment}}
\end{figure}
At rest, the drop remains essentially hemispherical since its radius ($R \sim$ 1.5 mm) does not exceed the capillary length ($l_c\sim 2.5 mm$ for water at room temperature) and thus surface tension overcomes gravity. The acoustic energy transmitted to the fluid induces zonal drop oscillations, of degree $l$=2 (in spherical-harmonics basis), and a tilt of the drop to the right, as shown in Fig.~\ref{fig:montage}. This left-right asymmetry is due to the asymmetry of the acoustic field, which is radiated in the drop according to a refraction angle $\phi_r \approx 25^o$ given by Snell-Descartes law, see Fig.~\ref{fig:idt}. When the drop is sufficiently tilted for the rear and front contact angles to exceed their hysteretic value, one additionally observes drop motion. The vertical amplitude $\Delta h$ and frequency $f_r$ of oscillation depend on both $f_m$ and $d$, see Fig.~\ref{fig:amplitudepcomment}. According to these parameters, either a harmonic response at frequency $f_r = f_m$, a subharmonic response at $f_r = 1/2 f_m$ or a superharmonic response at $f_r = 2 f_m$ is observed (see also accompanying movies in the supplemental material). 

\noindent\textbf{Harmonic region: shift of resonance.} In the blue area of Fig.~\ref{fig:amplitudepcomment}, a peak of response is obtained when $f_m$ reaches the inertio-capillary characteristic frequency $f_o$, which can be estimated from Rayleigh formula: $f_o = \left( 8 \gamma / 3 \pi \rho V \right)^{1/2} \approx 89$ Hz for the dipolar ($l=2$) oscillation of a $7.5 \mu l$ droplet, with $\gamma$ the surface tension and $V$ the droplet volume. At intermediate power ($d$=1.10 nm, $d$=1.38 nm), the peak is asymmetric, with a skewness directed to low frequencies and the resonance frequency decreases with the amplitude of oscillation.  This response is typical of an \textit{anharmonic} oscillator with softening spring ($\beta<0$):
\begin{equation}
\ddot{x} + 2 \lambda \dot{x} + \omega_o^2 x + \alpha x^2 + \beta x^3 = F cos (\Omega t)
\label{equation1}
\end{equation}
where $x$ is the dynamic variable (here the deformation of the drop), $t$ the time, $\lambda$ the damping coefficient, $\omega_o = 2 \pi f_o$ the angular eigen frequency, $\alpha$ and $\beta$ two nonlinearity coefficients, $F$ the amplitude of excitation and $\Omega$ the excitation frequency. Such nonlinear behavior has already been reported by Perez et al. \cite{pre_perez_2000} for levitating drops larger than the capillary length and more recently by Miyamoto et al. \cite{jjap_miyamoto_2002} for sessile droplets smaller than the capillary length, with pinned contact line. These authors determine the coefficient appearing in Eq.~(\ref{equation1}) from experiments and compare the frequency response of the drop to theoretical predictions. In our system the oscillation damping is due to dissipation in the viscous boundary layer and in the neighborhood of the contact line \cite{rmp_bonn_2009}. The nonlinear response of droplets appears when they undergo finite-amplitude deformations \cite{jfm_tsamopoulos_1983,jfm_smith_2010}, leading  to a shift of the resonance frequency to lower frequencies as the amplitude of oscillation increases. For sessile drop, additional nonlinearity results from the up-down asymmetry of boundaries and the presence of a contact line.

\noindent\textbf{Superharmonic region: combination of modes.} In the green region of Fig.~\ref{fig:amplitudepcomment}, low $f_m$ and large $\Delta h$, the drop response is a combination of harmonic and superharmonic modes. Indeed, the drop is pushed by the acoustic wave when the signal is on. Then, the drop keeps bouncing in the period with no forcing (signal is off) and oscillates a whole cycle before the next push. After a transient phase, this synchronization of forced and natural bouncing results in large drop oscillations. 

\noindent\textbf{Subharmonic region: parametric resonance.} In the red region of Fig.~\ref{fig:amplitudepcomment}, the droplet responds at $f_m / 2$. This subharmonic response appears only above a threshold: $d \ge$ 1.42 nm, with a frequency window broadening progressively with the amplitude of oscillation $\Delta h$. This is typical of a so-called \textit{Arnold Tongue}. Furthermore, the amplitude of the subharmonic response at fixed $f_m$ is independent of the amplitude of excitation $d$. All of these properties are characteristic of parametric resonance. Parametric instability of an oscillator is enabled when its characteristic frequency $\omega_o$ is modulated in time near $2 \omega_o$ \cite{ptrsl_faraday_1831}. It can be modeled with a Mathieu equation:
\begin{equation}
\ddot{x} + 2 \lambda \dot{x} + \omega_o^2 [1 + A \cos (2 \omega_o + \epsilon ) t]x
\end{equation}
with $\epsilon \ll \omega_o$. Such Mathieu equation can be obtained when an anharmonic oscillator described by Eq.~(\ref{equation1}) is excited near $2 \omega_o$: $\Omega = 2 \omega_o + \epsilon$ \cite{e_landau_1959_m}. Indeed, from the asymptotic expansion of the variable $x$ in Eq.~(\ref{equation1}): $x = x_1 + x_2$, with $x_1$ the solution of the harmonic oscillator and $x_2 \ll x_1$, we obtain at second order:
\begin{equation}
\ddot{x_2} + 2 \lambda \dot{x_2} + \omega_o^2 \left[1 - \frac{2 \alpha F}{3 w_o^4} \cos (2 \omega_o + \epsilon) t + ... \right] = 0
\end{equation}
The parametric excitation of \textit{zonal} oscillation modes was predicted theoretically by Papoular and Parayre \cite{prl_papoular_1997} for levitating drops but not observed experimentally. Indeed, parametric resonance appears for anharmonic oscillators above an amplitude threshold $F_t = 6 \lambda \omega_o^3 / | \alpha |$  which decreases with the nonlinearity coefficient $\alpha$. This latter is related to the asymmetry of the oscillator stiffness for prolate and oblate deformation ($x<0$ or $>0$) which is increased by the presence of the substrate. Thus, the observation of the parametric response in the present experiments could be explained by a larger nonlinearity of the system and larger excitations than in the system of Papoular and Parayre.
\begin{figure}[htbp]
\includegraphics[width=0.4\textwidth]{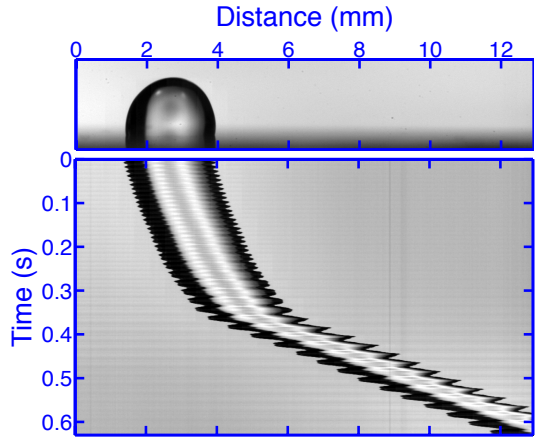}
\caption{(color online) Evolution of a 7.5 $\mu$l droplet excited by an acoustic wave of amplitude $d$ = 1.55 nm and modulation frequency $f_m$ = 100 Hz. The top picture shows the initial shape of the drop. The spatio-temporal diagram below is obtained by taking the base line of the drop (dashed blue line on top picture) and showing its evolution as a function of time. The average slope of the front or rear curve delimiting the drop gives the velocity of the drop. \label{fig:reslicepdroplet}}
\end{figure}

The parametric mode appears after a transient phase, as shown in the spatio-temporal diagram on Fig.~\ref{fig:reslicepdroplet}. At first, the droplet oscillates at the same frequency as the excitation $f_m$ = 100 Hz between t = 0 and 0.15 s. Between t = 0.15 and t = 0.32 s, the parametric instability grows. Finally, the droplet reaches a stationary regime of oscillation at half the frequency of modulation $f_r = 1/2 f_m$. Between initial harmonic and later parametric response, the amplitude of droplet lateral oscillations $\Delta w$ is increased by a factor of 4.8. This large increase of $\Delta w$ comes along with an increase of the droplet velocity by a factor of 4, see the rupture of slope at t = 0.35 s in Fig. \ref{fig:reslicepdroplet}.

\begin{figure}[htbp]
\includegraphics[width=0.5\textwidth]{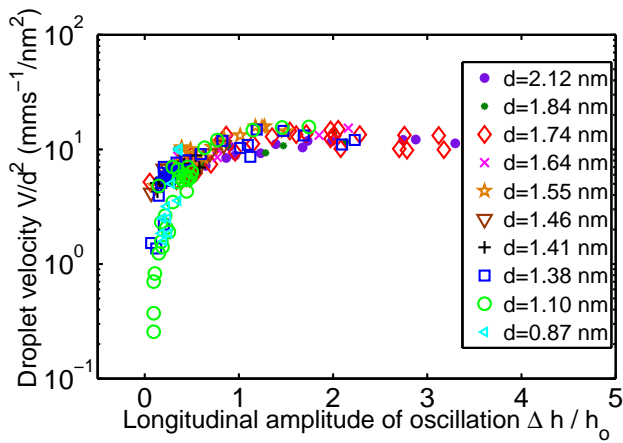}
\caption{(color online) Velocity of the drop divided by the square of the surface acoustic wave amplitude $d^2$ (corresponding to the acoustic power radiated into the drop up to a prefactor) as a function of the longitudinal amplitude of oscillation $\Delta h / h_o$. Each marker corresponds to a specific amplitude $d$.\label{fig:velocityvsamplitudelaterale}}
\end{figure}

In all the previously described experiments, there is indeed a correlation between the droplets velocity $V$ and their amplitude of oscillation $\Delta h / h_o$, see Fig.~\ref{fig:velocityvsamplitudelaterale}. Indeed, the date of the drop velocity divided by the applied acoustic power collapse into a single curve, which increases with the amplitude of oscillation up to $\Delta h / h_o= 1$, and then reaches a plateau. From this figure, we can conclude that (i) the droplet velocity is basically proportional to the surface acoustic wave power and (ii) that droplets oscillations promote their mobility.
Indeed, the velocity of a drop is determined by the equilibrium between the driving forces and the retention forces. The motion is due to the contact angle difference between the front and rear interfaces and thus, to the asymmetry of the drop, see e.g. Ref. \onlinecite{rmp_bonn_2009}. An increase of the acoustic wave amplitude $d$ naturally leads to larger droplet deformations and therefore larger asymmetry and velocities of the drop. The signal modulation does not affect the acoustic power. Nevertheless, it can affect both the driving and retention forces into different ways. First, it is important to mention that while some authors, see e.g. Ref. \onlinecite{l_decker_1996}, have observed a decrease of the hysteresis of the contact line due to its continuous depinning, such variations are not observed in the present study and cannot explain the increase of drop velocity. Second, as seen previously, the droplet \textit{static} shape is not much affected by gravity since the droplet radius is smaller than the capillary length. However, when the drop is highly stretched with a left/right asymmetry component (see e.g. Fig. 1b), gravity plays a significant role during the retraction phase. The drop is not simply pulled back by capillary forces, but also the upper part of the liquid drop falls vertically by gravity (see accompanying movie ``superharmonic'' in supplemental material): this strengthens even further the left/right asymmetry and should contribute to the drop motion. Finally, the acoustic energy transferred to the drop and the retention force depend respectively on the contact surface with the substrate and the perimeter of the contact line. Thus a stretching of the drop contributes to a reduction of the retention force but also a reduction of the amount of energy transferred to the drop. This might explain the existence of the plateau region in Fig. \ref{fig:velocityvsamplitudelaterale}. To conclude, we can compare the velocities obtained with and without signal modulation at same acoustic power (see Fig. \ref{fig:dvelocityvsfrequency} and accompanying movies in the supplemental material). The drop velocity can be increased by a factor 100 thanks to the modulation.

\begin{figure}[htbp]
\includegraphics[width=0.5\textwidth]{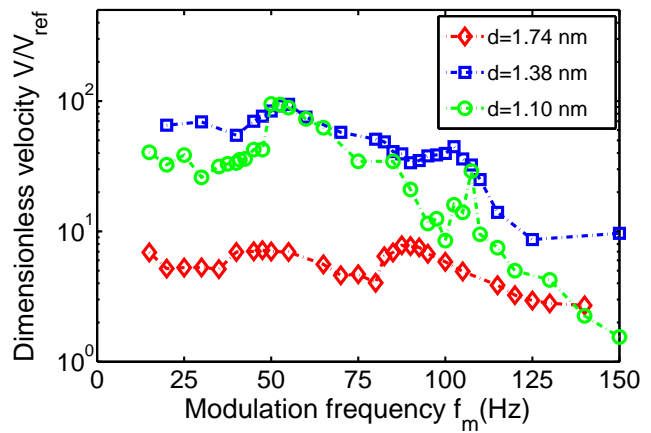}
\caption{(color online) Velocity of the drop V measured with a modulated signal divided by a reference velocity $V_{ref}$ obtained at same acoustic power when the modulation is turned off, as a function of the modulation frequency $f_m$.  \label{fig:dvelocityvsfrequency}}
\end{figure}

 In this paper, we have shown that inertio-capillary modes of oscillations can be excited either directly or parametrically by modulating the acoustic signal around once or twice the Rayleigh-Lamb characteristic frequency. This modulation introduces an original way to decrease the acoustic power required to stir inside, to stretch or to move a sessile droplet with SAWs. Compared to the non-modulated case, the minimum acoustic power required to move a droplet at a non-zero speed, and a speed of 5 mm s$^{-1}$ are reduced by a factor of 2 and 3 respectively, while the minimum power required to stretch the droplet vertically by a factor $\Delta h / h_o$ of 0.2 and 1 are reduced by a factor 5 and 3 respectively. This is especially important for the harmless manipulations of biofluids which could be damaged by the temperature increase due to the dissipation of acoustic energy. It is also of primary interest for portable lab-on-a-chip which requires low power consumption. \\

\bibliography{biblioa,bibliob}

\begin{thebibliography}{25}
\expandafter\ifx\csname natexlab\endcsname\relax\def\natexlab#1{#1}\fi
\expandafter\ifx\csname bibnamefont\endcsname\relax
  \def\bibnamefont#1{#1}\fi
\expandafter\ifx\csname bibfnamefont\endcsname\relax
  \def\bibfnamefont#1{#1}\fi
\expandafter\ifx\csname citenamefont\endcsname\relax
  \def\citenamefont#1{#1}\fi
\expandafter\ifx\csname url\endcsname\relax
  \def\url#1{\texttt{#1}}\fi
\expandafter\ifx\csname urlprefix\endcsname\relax\def\urlprefix{URL }\fi
\providecommand{\bibinfo}[2]{#2}
\providecommand{\eprint}[2][]{\url{#2}}

\bibitem[{\citenamefont{Stone et~al.}(2004)\citenamefont{Stone, Stroock, and
  Adjari}}]{arfm_stone_2004}
\bibinfo{author}{\bibfnamefont{H.}~\bibnamefont{Stone}},
  \bibinfo{author}{\bibfnamefont{A.}~\bibnamefont{Stroock}}, \bibnamefont{and}
  \bibinfo{author}{\bibfnamefont{A.}~\bibnamefont{Adjari}},
  \bibinfo{journal}{Ann. Rev. Fluid Mech.} \textbf{\bibinfo{volume}{36}},
  \bibinfo{pages}{381} (\bibinfo{year}{2004}).

\bibitem[{\citenamefont{Pollack et~al.}(2000)\citenamefont{Pollack, Fair, and
  Shenderov}}]{apl_pollack_2000}
\bibinfo{author}{\bibfnamefont{M.}~\bibnamefont{Pollack}},
  \bibinfo{author}{\bibfnamefont{R.}~\bibnamefont{Fair}}, \bibnamefont{and}
  \bibinfo{author}{\bibfnamefont{A.}~\bibnamefont{Shenderov}},
  \bibinfo{journal}{Appl. Phys. Lett.} \textbf{\bibinfo{volume}{77}},
  \bibinfo{pages}{1725} (\bibinfo{year}{2000}).

\bibitem[{\citenamefont{Baroud et~al.}(2007)\citenamefont{Baroud, Robert~de
  Saint~Vincent, and Delville}}]{loc_baroud_2007}
\bibinfo{author}{\bibfnamefont{C.}~\bibnamefont{Baroud}},
  \bibinfo{author}{\bibfnamefont{M.}~\bibnamefont{Robert~de Saint~Vincent}},
  \bibnamefont{and} \bibinfo{author}{\bibfnamefont{J.}~\bibnamefont{Delville}},
  \bibinfo{journal}{Lab on a chip} \textbf{\bibinfo{volume}{7}},
  \bibinfo{pages}{1029} (\bibinfo{year}{2007}).

\bibitem[{\citenamefont{Friend and Yeo}(2011)}]{rmp_friend_2011}
\bibinfo{author}{\bibfnamefont{J.}~\bibnamefont{Friend}} \bibnamefont{and}
  \bibinfo{author}{\bibfnamefont{L.}~\bibnamefont{Yeo}}, \bibinfo{journal}{Rev.
  Mod. Phys.} \textbf{\bibinfo{volume}{83}}, \bibinfo{pages}{647}
  (\bibinfo{year}{2011}).

\bibitem[{\citenamefont{Kondoh et~al.}(2009)\citenamefont{Kondoh, Shimizu,
  Matsui, Sugimoto, and Shiokawa}}]{kondoh_saa_2009}
\bibinfo{author}{\bibfnamefont{J.}~\bibnamefont{Kondoh}},
  \bibinfo{author}{\bibfnamefont{N.}~\bibnamefont{Shimizu}},
  \bibinfo{author}{\bibfnamefont{Y.}~\bibnamefont{Matsui}},
  \bibinfo{author}{\bibfnamefont{M.}~\bibnamefont{Sugimoto}}, \bibnamefont{and}
  \bibinfo{author}{\bibfnamefont{S.}~\bibnamefont{Shiokawa}},
  \bibinfo{journal}{Sensors and actuators A} \textbf{\bibinfo{volume}{149}},
  \bibinfo{pages}{292} (\bibinfo{year}{2009}).

\bibitem[{\citenamefont{Beyssen}(2006)}]{phd_beyssen_2006}
\bibinfo{author}{\bibfnamefont{D.}~\bibnamefont{Beyssen}}, Ph.D. thesis,
  \bibinfo{school}{Universit\'{e} Henri Poincar\'{e}} (\bibinfo{year}{2006}).

\bibitem[{\citenamefont{Renaudin et~al.}(2010)\citenamefont{Renaudin, Chabot,
  Grondin, Aimez, and Charette}}]{loc_renaudin_2010}
\bibinfo{author}{\bibfnamefont{A.}~\bibnamefont{Renaudin}},
  \bibinfo{author}{\bibfnamefont{V.}~\bibnamefont{Chabot}},
  \bibinfo{author}{\bibfnamefont{E.}~\bibnamefont{Grondin}},
  \bibinfo{author}{\bibfnamefont{V.}~\bibnamefont{Aimez}}, \bibnamefont{and}
  \bibinfo{author}{\bibfnamefont{P.}~\bibnamefont{Charette}},
  \bibinfo{journal}{Lab Chip} \textbf{\bibinfo{volume}{10}},
  \bibinfo{pages}{111} (\bibinfo{year}{2010}).

\bibitem[{\citenamefont{Tan et~al.}(2009)\citenamefont{Tan, Friend, and
  Yeo}}]{prl_tan_2009}
\bibinfo{author}{\bibfnamefont{M.}~\bibnamefont{Tan}},
  \bibinfo{author}{\bibfnamefont{J.}~\bibnamefont{Friend}}, \bibnamefont{and}
  \bibinfo{author}{\bibfnamefont{L.}~\bibnamefont{Yeo}},
  \bibinfo{journal}{Phys. Rev. Lett.} \textbf{\bibinfo{volume}{103}},
  \bibinfo{pages}{024501} (\bibinfo{year}{2009}).

\bibitem[{\citenamefont{Wixforth}(2003)}]{sm_wixforth_2003}
\bibinfo{author}{\bibfnamefont{A.}~\bibnamefont{Wixforth}},
  \bibinfo{journal}{Superlattices and Microstructures}
  \textbf{\bibinfo{volume}{33}}, \bibinfo{pages}{389} (\bibinfo{year}{2003}).

\bibitem[{\citenamefont{Rayleigh}(1879)}]{prsl_rayleigh_1879}
\bibinfo{author}{\bibfnamefont{J.}~\bibnamefont{Rayleigh}},
  \bibinfo{journal}{Proc. R. Soc. Lond.} \textbf{\bibinfo{volume}{29}},
  \bibinfo{pages}{71} (\bibinfo{year}{1879}).

\bibitem[{\citenamefont{Lamb}(1932)}]{cup_lamb_1932}
\bibinfo{author}{\bibfnamefont{H.}~\bibnamefont{Lamb}},
  \emph{\bibinfo{title}{Hydrodynamics}} (\bibinfo{publisher}{Cambridge
  University Press}, \bibinfo{address}{England}, \bibinfo{year}{1932}).

\bibitem[{\citenamefont{Strani and Sabetta}(1984)}]{jfm_strani_1984}
\bibinfo{author}{\bibfnamefont{M.}~\bibnamefont{Strani}} \bibnamefont{and}
  \bibinfo{author}{\bibfnamefont{F.}~\bibnamefont{Sabetta}},
  \bibinfo{journal}{J. Fluid Mech.} \textbf{\bibinfo{volume}{141}},
  \bibinfo{pages}{233} (\bibinfo{year}{1984}).

\bibitem[{\citenamefont{Noblin et~al.}(2004)\citenamefont{Noblin, Buguin, and
  Brochard-Wyart}}]{epje_noblin_2004}
\bibinfo{author}{\bibfnamefont{X.}~\bibnamefont{Noblin}},
  \bibinfo{author}{\bibfnamefont{A.}~\bibnamefont{Buguin}}, \bibnamefont{and}
  \bibinfo{author}{\bibfnamefont{F.}~\bibnamefont{Brochard-Wyart}},
  \bibinfo{journal}{Eur. Phys. J. E.} \textbf{\bibinfo{volume}{14}},
  \bibinfo{pages}{395} (\bibinfo{year}{2004}).

\bibitem[{\citenamefont{Yamakita et~al.}(1999)\citenamefont{Yamakita, Matsui,
  and Shiokawa}}]{jjap_yamakita_1999}
\bibinfo{author}{\bibfnamefont{S.}~\bibnamefont{Yamakita}},
  \bibinfo{author}{\bibfnamefont{Y.}~\bibnamefont{Matsui}}, \bibnamefont{and}
  \bibinfo{author}{\bibfnamefont{S.}~\bibnamefont{Shiokawa}},
  \bibinfo{journal}{Jpn. J; Appl. Phys} \textbf{\bibinfo{volume}{38}},
  \bibinfo{pages}{3127} (\bibinfo{year}{1999}).

\bibitem[{\citenamefont{Brunet et~al.}(2010)\citenamefont{Brunet, Baudoin,
  Matar, and Zoueshtiagh}}]{pre_brunet_2010}
\bibinfo{author}{\bibfnamefont{P.}~\bibnamefont{Brunet}},
  \bibinfo{author}{\bibfnamefont{M.}~\bibnamefont{Baudoin}},
  \bibinfo{author}{\bibfnamefont{O.}~\bibnamefont{Matar}}, \bibnamefont{and}
  \bibinfo{author}{\bibfnamefont{F.}~\bibnamefont{Zoueshtiagh}},
  \bibinfo{journal}{Phys. Rev. E} \textbf{\bibinfo{volume}{81}},
  \bibinfo{pages}{036315} (\bibinfo{year}{2010}).

\bibitem[{\citenamefont{Alzuaga et~al.}(2005)\citenamefont{Alzuaga, Manceau,
  and Bastien}}]{jsv_alzuaga_2005}
\bibinfo{author}{\bibfnamefont{S.}~\bibnamefont{Alzuaga}},
  \bibinfo{author}{\bibfnamefont{J.}~\bibnamefont{Manceau}}, \bibnamefont{and}
  \bibinfo{author}{\bibfnamefont{F.}~\bibnamefont{Bastien}},
  \bibinfo{journal}{J. Sound Vib.} \textbf{\bibinfo{volume}{282}},
  \bibinfo{pages}{151} (\bibinfo{year}{2005}).

\bibitem[{\citenamefont{Papoular and Parayre}(1997)}]{prl_papoular_1997}
\bibinfo{author}{\bibfnamefont{M.}~\bibnamefont{Papoular}} \bibnamefont{and}
  \bibinfo{author}{\bibfnamefont{C.}~\bibnamefont{Parayre}},
  \bibinfo{journal}{Phys. Rev. Lett.} \textbf{\bibinfo{volume}{78}},
  \bibinfo{pages}{2120} (\bibinfo{year}{1997}).

\bibitem[{\citenamefont{Perez et~al.}(2000)\citenamefont{Perez, Salvo,
  Su\'{e}ry, Br\'{e}chet, and Papoular}}]{pre_perez_2000}
\bibinfo{author}{\bibfnamefont{M.}~\bibnamefont{Perez}},
  \bibinfo{author}{\bibfnamefont{L.}~\bibnamefont{Salvo}},
  \bibinfo{author}{\bibfnamefont{M.}~\bibnamefont{Su\'{e}ry}},
  \bibinfo{author}{\bibfnamefont{Y.}~\bibnamefont{Br\'{e}chet}},
  \bibnamefont{and} \bibinfo{author}{\bibfnamefont{M.}~\bibnamefont{Papoular}},
  \bibinfo{journal}{Phys. Rev. E} \textbf{\bibinfo{volume}{61}},
  \bibinfo{pages}{2669} (\bibinfo{year}{2000}).

\bibitem[{\citenamefont{Miyamoto et~al.}(2002)\citenamefont{Miyamoto, Nagamoto,
  Matsui, and Shiokawa}}]{jjap_miyamoto_2002}
\bibinfo{author}{\bibfnamefont{K.}~\bibnamefont{Miyamoto}},
  \bibinfo{author}{\bibfnamefont{S.}~\bibnamefont{Nagamoto}},
  \bibinfo{author}{\bibfnamefont{Y.}~\bibnamefont{Matsui}}, \bibnamefont{and}
  \bibinfo{author}{\bibfnamefont{S.}~\bibnamefont{Shiokawa}},
  \bibinfo{journal}{Jpn. J. Appl. Phys.} \textbf{\bibinfo{volume}{41}},
  \bibinfo{pages}{3465} (\bibinfo{year}{2002}).

\bibitem[{\citenamefont{Bonn et~al.}(2009)\citenamefont{Bonn, Eggers, Indekeu,
  Meunier, and Rolley}}]{rmp_bonn_2009}
\bibinfo{author}{\bibfnamefont{D.}~\bibnamefont{Bonn}},
  \bibinfo{author}{\bibfnamefont{J.}~\bibnamefont{Eggers}},
  \bibinfo{author}{\bibfnamefont{J.}~\bibnamefont{Indekeu}},
  \bibinfo{author}{\bibfnamefont{J.}~\bibnamefont{Meunier}}, \bibnamefont{and}
  \bibinfo{author}{\bibfnamefont{E.}~\bibnamefont{Rolley}},
  \bibinfo{journal}{Rev. Mod. Phys.} \textbf{\bibinfo{volume}{81}},
  \bibinfo{pages}{739} (\bibinfo{year}{2009}).

\bibitem[{\citenamefont{Tsamopoulos and Brown}(1983)}]{jfm_tsamopoulos_1983}
\bibinfo{author}{\bibfnamefont{R.}~\bibnamefont{Tsamopoulos}} \bibnamefont{and}
  \bibinfo{author}{\bibfnamefont{R.}~\bibnamefont{Brown}}, \bibinfo{journal}{J.
  Fluid Mech.} \textbf{\bibinfo{volume}{127}}, \bibinfo{pages}{519}
  (\bibinfo{year}{1983}).

\bibitem[{\citenamefont{Smith}(2010)}]{jfm_smith_2010}
\bibinfo{author}{\bibfnamefont{W.}~\bibnamefont{Smith}}, \bibinfo{journal}{J.
  Fluid Mech.} \textbf{\bibinfo{volume}{654}}, \bibinfo{pages}{141}
  (\bibinfo{year}{2010}).

\bibitem[{\citenamefont{Faraday}(1831)}]{ptrsl_faraday_1831}
\bibinfo{author}{\bibfnamefont{M.}~\bibnamefont{Faraday}},
  \bibinfo{journal}{Phil. Trans. R. Soc.} \textbf{\bibinfo{volume}{121}},
  \bibinfo{pages}{299} (\bibinfo{year}{1831}).

\bibitem[{\citenamefont{Landau and Lifshitz}(1959)}]{e_landau_1959_m}
\bibinfo{author}{\bibfnamefont{L.}~\bibnamefont{Landau}} \bibnamefont{and}
  \bibinfo{author}{\bibnamefont{Lifshitz}}, \emph{\bibinfo{title}{Course of
  theoretical Physics, Mechanics}} (\bibinfo{publisher}{Pergamon Press, New
  York}, \bibinfo{year}{1959}).

\bibitem[{\citenamefont{Decker and Garoff}(1996)}]{l_decker_1996}
\bibinfo{author}{\bibfnamefont{E.}~\bibnamefont{Decker}} \bibnamefont{and}
  \bibinfo{author}{\bibfnamefont{S.}~\bibnamefont{Garoff}},
  \bibinfo{journal}{Langmuir} \textbf{\bibinfo{volume}{12}},
  \bibinfo{pages}{2100} (\bibinfo{year}{1996}).

\end{thebibliography}

\end{document}